\documentclass[lettersize,journal]{IEEEtran}
\usepackage{amsmath,amsfonts}
\usepackage{algorithmic}
\usepackage{algorithm}
\usepackage{array}
\usepackage{textcomp}
\usepackage{stfloats}
\usepackage{url}
\usepackage{verbatim}
\usepackage{graphicx}
\usepackage{cite}
\usepackage{enumitem}
\usepackage{breqn}
\usepackage[caption=false,font=footnotesize,labelfont=sf,textfont=rm]{subfig}
\usepackage{times}

\begin{document}

\title{Embroidery Actuator Utilizing Embroidery Patterns \\to Generate Diverse Fabric Deformations}

\author{Yuki OTA (Nagoya Univ.), Yuki FUNABORA(Nagoya Univ.)
        
\thanks{This paper was supported by JST FOREST Program (Grant Number JPMJFR216T).
}
}

\maketitle

\begin{abstract}
This paper presents a novel Embroidery Actuator, a fabric-integrated pneumatic actuator that enables diverse and controllable deformations through embroidery pattern design. Unlike conventional fabric actuators that rely on fiber- or thread-shaped actuators, the proposed actuator is fabricated by directly stitching an inflatable tube onto the fabric using a cord-embroidery technique. The embroidered thread and the fabric jointly form a sleeve that constrains the expansion of the inflatable tube, converting internal pressure into targeted bending or stretching deformations.
By varying the embroidery pattern, such as zigzag or cross configurations, different geometric constraints can be realized, allowing for flexible control of deformation direction and magnitude. Analytical deformation models based on the \textit{Neo-Hookean model} and \textit{Lagrange's equations} were developed to predict the relationship between pneumatic pressure and bending angle, and were experimentally validated using motion-capture measurements. The results demonstrated that the actuator achieves strong agreement with the analytical deformation model. 
\end{abstract}


\section{Introduction}
\IEEEPARstart{S}{oft} actuators have attracted significant attention in recent years due to their inherent flexibility, low weight, and safe physical interaction with humans \cite{Review1,Review2}.
These characteristics make them ideal for applications in wearable, medical, and rehabilitation devices, where compliance and comfort are essential \cite{wear1,wear2,wear3}. 
Among various types of soft actuators, fabric-based actuators are of particular interest \cite{NunoReview1,NunoReview2,NunoReview3}. They can conform to the body surface and generate distributed forces suitable for massage, haptic feedback, or assistive motion \cite{Zendou, Haptiknit, SMAPower}.

Conventional fabric-type actuators are commonly constructed by integrating or knitting thread-type actuators, such as shape-memory alloy fibers or McKibben artificial muscles, into textile structures. 
Although such designs benefit from well-established deformation models of fiber actuators, their achievable deformations are limited by the mechanical properties of the embedded threads and the manner in which they are fixed to the fabric. 
These constraints often restrict deformation to simple bending or contraction modes. 
Other studies have proposed fabric actuators based on inflatable rubber tubes sandwiched between textile layers; however, these designs primarily induce one-dimensional stretching and offer limited versatility in shaping complex, multidirectional deformations.

To overcome these limitations, we propose the Embroidery Actuator—a new class of fabric actuator that does not rely on fiber-shaped actuators. The actuator is fabricated by directly stitching an inflatable tube onto the surface of a fabric using a cord-embroidery technique. In this configuration, the embroidery and the fabric together form an external sleeve that constrains and guides the inflation of the tube. The resulting deformation depends strongly on the embroidery pattern, which can be freely varied by design. This allows for a wide range of deformation modes to be realized with ease. Furthermore, because the fabric, tube, and thread can be independently chosen, the actuator’s size, force, and operating pressure can be scaled and customized for various applications.
\IEEEpubidadjcol

The main contributions of this paper are summarized as follows:

\begin{itemize}
\item A fabrication method for the Embroidery Actuator using a fiber-sewing machine.
\item Development and evaluation of the analytical deformation models of the Embroidery Actuator.
\item Experimental investigation of the deformation behavior under different embroidery patterns.
\end{itemize}

The remainder of this paper is organized as follows. Section~\ref{Sec_Related} reviews related studies on fiber- and fabric-type actuators and highlights the contributions of the proposed method. Section~\ref{Sec_Fabrication} describes the design and fabrication process of the Embroidery Actuator. Section~\ref{Sec_Model} presents an analytical deformation model, and Section~\ref{Sec_Exp} and Section~\ref{Sec_Result} experimentally validate its predictions. Finally, Section~\ref{Sec_Conclusion} concludes the paper and discusses potential directions for future work.

\section{Related Works}

\label{Sec_Related}

Various types of actuators have been developed for applications in industrial, medical, and other fields.
The Embroidery Actuator builds upon prior studies of thread-type and fabric-type actuators.
Therefore, this section introduces those previous works and clarifies the contributions of the Embroidery Actuator through comparison.

\subsection{Thread-Type Actuators}
Many thread-type actuators utilizing fluidic pressure have also been developed. McKibben artificial muscle is a representative example.
In the McKibben artificial muscle, the expansion generated by pneumatic pressurization of an internal inflatable tube is constrained by a braided sleeve, converting the radial expansion into axial contraction through the pantograph principle \cite{MckibbenMode2000}.
Beyond conventional contraction, Krishnan et al. achieved axial extension and torsion by altering the braiding angle and the balance of fiber orientations wound in opposite directions \cite{senikyouka}.
Furthermore, Ozgun et al. developed the Omnifiber, which enables extension, bending, and twisting motions by integrating additional fibers or plates into the braided sleeve \cite{Omnifiber}.
Likewise, thread-type actuators have been extensively developed using various materials such as shape-memory alloys (SMA) and liquid-crystal elastomers (LCE)\cite{SMAReview,LCEReview}.

\subsection{Fabric-Type Actuators}

By integrating or knitting these thread-type actuators into textiles, various fabric-type actuators have been studied and developed.
Hiramatsu et al. developed deformable fabrics by weaving cords with thin McKibben artificial muscles \cite{ThinMcKibben, Activetextile1, Activetextile2}.
Our research group has also proposed methods for fabricating fabric actuators by attaching thin McKibben actuators directly to textile surfaces \cite{nunoaku1,nunoaku2,nakagawa}.
In these approaches using thin artificial muscles, diverse fabric-surface deformations can be achieved by varying the weaving pattern and spatial arrangement of the muscles.
However, the achievable deformations are limited to bending based on the contraction of the artificial muscles, imposing restrictions on the range of deformation modes and actuator layouts.
Afsar et al. fabricated a fabric actuator capable of bidirectional stretching and pushing/pulling motion by incorporating Omnifiber into a textile \cite{OmniCorsetto}.
Nevertheless, the deformation remained limited to in-plane contraction of the fabric.
Several research groups have fabricated fabric actuators by weaving SMA fibers into fabrics \cite{SMAPower,SMAGlove,SMA}.
However, SMA-based actuators suffer from long response times and raise concerns regarding heat generation, potential effects on the human body, and discomfort during wear.
LCE fibers have also been explored for integration into fabric-based actuators, but their response speed and output force are limited \cite{LCEnuno1,LCEnuno2,LCEnuno3}.

These fabric-type actuators have the advantage of being able to exploit established deformation models of thread-type actuators; however, the overall deformation of the fabric is constrained by the mechanical properties of the thread-type actuators themselves.
Moreover, interactions between the fabric and fibers vary significantly depending on the integration and fixation methods, which affect the deformation of the fabric.
As an actuator that does not rely on thread-type actuators, Zhu et al. proposed the Fluidic Fabric Muscle Sheet (FFMS) \cite{FFMS}.
The FFMS is a fabric-integrated actuator in which a rubber tube is sandwiched between fabric layers.
Its circumferential expansion is constrained by the fabric, converting the internal pressure into axial stretching and bending motion.
While the FFMS directly generates fabric deformation from the expansion of the rubber tube, its deformation is essentially limited to simple one-dimensional extension and contraction.

Our proposed Embroidery Actuator is a fabric-integrated actuator constructed by stitching an inflatable tube onto the fabric using a cord-embroidery technique.
The embroidery thread and the base fabric together form an external sleeve that constrains and directs the inflation, converting it into targeted fabric-surface deformation.
By leveraging embroidery techniques, diverse sleeve structures can be readily constructed, enabling a wide range of complex deformations to be realized with ease.

\section{Embroidery Actuator}

\label{Sec_Fabrication}

This section describes the design, operating principle, and fabrication method of the Embroidery Actuator.

\subsection{Design and Principle}

\begin{figure}[t]
\centering
\includegraphics[width=1.0\linewidth]{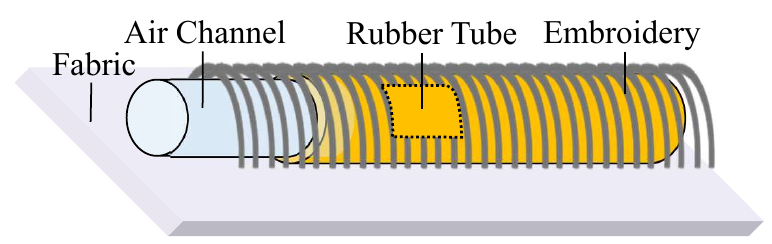}
\caption{Structure of the Embroidery Actuator.The threads and the fabric jointly enclose the inflatable tube, to which the air channel is connected.}
\label{Fig_Structure}
\end{figure}

The Embroidery Actuator has a hierarchical structure, as illustrated in Fig.~\ref{Fig_Structure}(a).
An inflatable tube is stitched onto the fabric using embroidery threads so that the threads and fabric jointly enclose the tube.
The actuator operates when pneumatic pressure is applied to the internal inflatable tube.
The mechanical constraint applied to the tube is non-uniform: it differs between the embroidery-thread side and the fabric side.
In particular, the constraint from the embroidery side varies depending on the embroidery pattern.
Therefore, by altering the embroidery pattern, the degree and direction of constraint on the tube’s expansion can be controlled, thereby introducing anisotropic expansion and enabling diverse deformation behaviors.

\subsection{Fabrication}

The Embroidery Actuator primarily consists of an inflatable tube, embroidery thread, and fabric.
The deformation characteristics depend on the material parameters such as elasticity and stiffness of these components.
In this study, the actuator was fabricated using a natural-rubber inflatable tube (Hagitec, T-2 02-089-01-02), polyester sewing thread, and cotton fabric as the substrate.

\begin{figure*}
\centering
\includegraphics[width=1.0\linewidth]{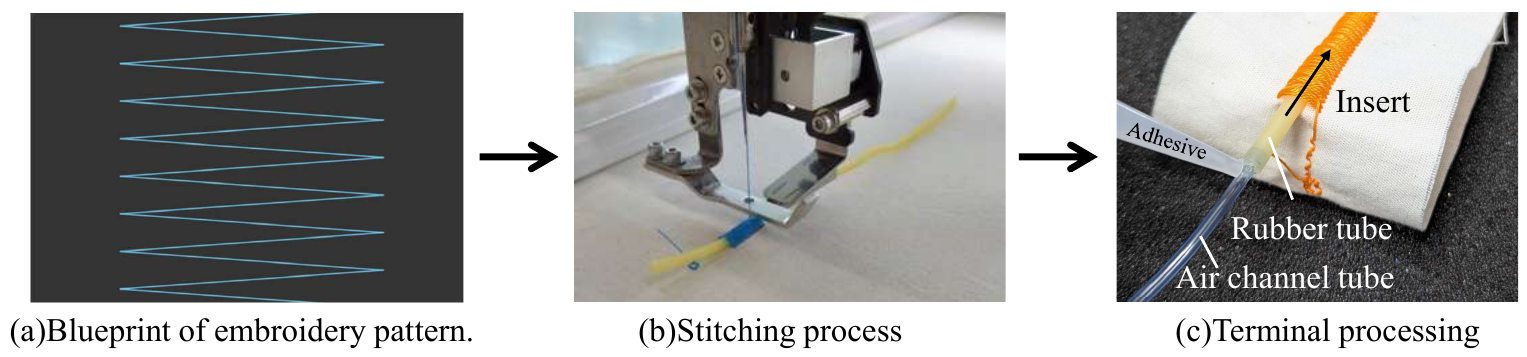}
\caption{Fabrication process of the Embroidery Actuator: (a)Designing the embroidery pattern that stitches the inflatable tube onto the fabric.(b) Stitching inflatable tube onto the fabric.(c) Processing the terminals of the inflatable tube to enable pneumatic pressurization.}
\label{Fig_Fabrication}
\end{figure*}

The fabrication process of the Embroidery Actuator consists of three main stages:
(a) embroidery design, (b) stitching process, and (c) end-terminal processing, as shown in Fig.~\ref{Fig_Fabrication}.
The details are described below.

\paragraph{Embroidery Design}

\begin{figure}[tb]
    \centering
    \includegraphics[width=1\linewidth]{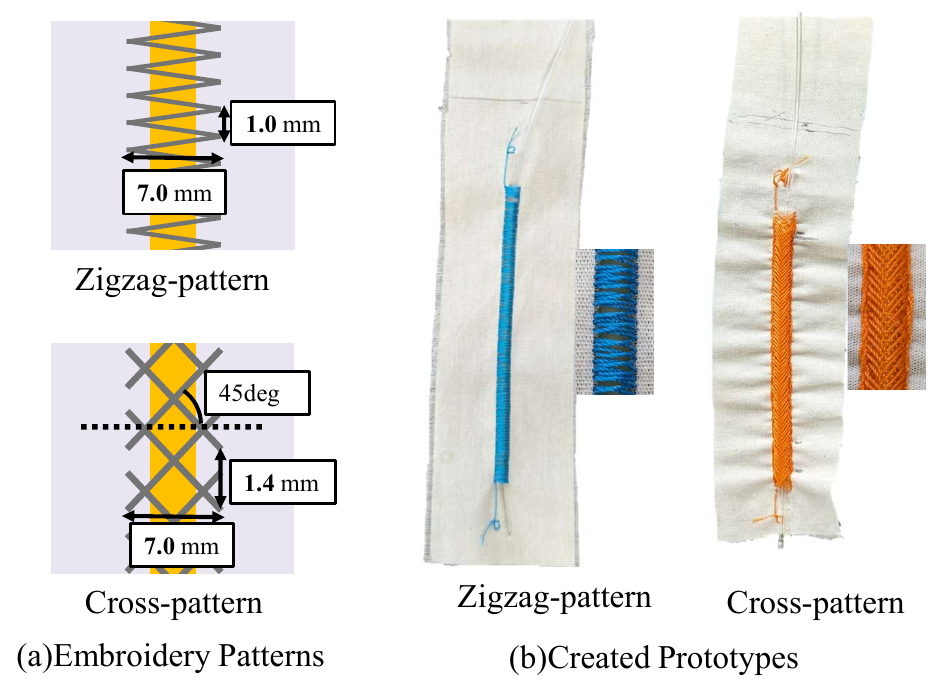}
    \caption{Embroidery patterns and photographs of the prototypes: (a)For the zigzag-pattern actuators, the stitch interval was set to 1.0~mm, and embroidery width was 7~mm. For the cross-pattern actuators, the stitch interval was set to 1.4~mm, embroidery width was 7~mm, and embroidery angle is 45~deg.(b)The fabricated prototypes.}
    \label{Fig_Prototype}
\end{figure}

In this stage, the embroidery pattern that stitches the inflatable tube onto the fabric surface is designed.
The deformation direction and magnitude of the Embroidery Actuator depend on the design parameters of the embroidery pattern, including pattern type, embroidery width, embroidery angle, and spacing between adjacent patterns.
In this work, embroidery design software (TISM DG16) was used to design two types of actuators with zigzag and cross-patterns.
The designed patterns and corresponding parameters are illustrated in Fig.~\ref{Fig_Prototype}(a).

\paragraph{Stitching Process}

Next, the inflatable tube is stitched onto the fabric according to the designed embroidery pattern.
The embroidery threads are sewn so that they cross over and hold the tube placed on the fabric surface.
The stitching process was performed using a composite fiber-sewing machine (TISM, TCWM-101).
By setting the inflatable tube, thread, and fabric on the machine, the stitching is carried out automatically following the pre-defined embroidery pattern.

\paragraph{Terminal Processing}

Finally, the terminals of the inflatable tube are processed to enable pneumatic pressurization.
This involves sealing the tube ends and preparing the air passage, as follows:
First, the inflatable tube was cut so that its ends did not protrude beyond the embroidered region.
Then, polyurethane tubes ($\phi$2 mm, Koganei U2-C-100) were inserted into both ends of the inflatable tube and sealed completely with adhesive.
One of the inserted polyurethane tubes served as the air-supply passage, while the other was used as a sealed terminal.
The terminal side was cut to a length of 30 mm, filled with adhesive, and hardened to ensure airtight sealing.

\begin{figure}[tb]
        \centering
        \subfloat[Zigzag-pattern]{\includegraphics[width=0.45\linewidth]{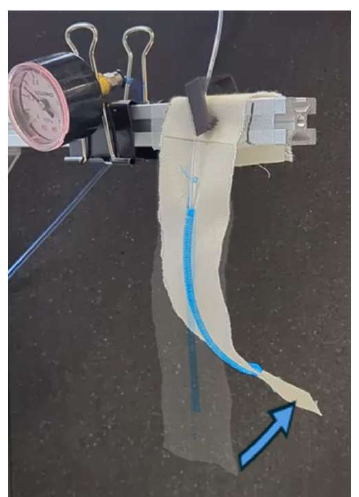}}
        \subfloat[Cross-pattern]{\includegraphics[width=0.45\linewidth]{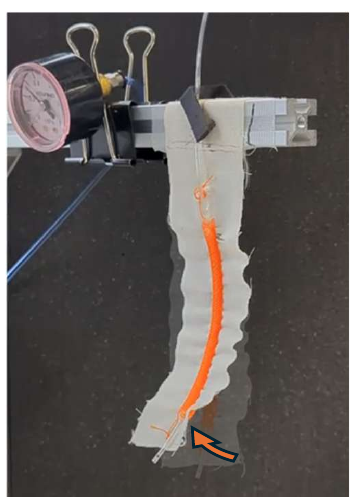}}
        \caption{Deformation of prototypes: (a)Bending toward the front side of the fabric.(b)Bending toward the back side of the fabric.}
        \label{Fig_Deformation}
\end{figure}

The fabricated Embroidery Actuator is shown in Fig.~\ref{Fig_Prototype}(b), and its deformation behavior is presented in Fig.~\ref{Fig_Deformation}.
The actuator with the zigzag-pattern bent toward the back side of the fabric, whereas in the cross-pattern, it bent toward the front side.

\section{Modeling}

\label{Sec_Model}

This section derives the analytical deformation model of the Embroidery Actuator under applied pneumatic pressure.

In the Embroidery Actuator, the sleeve does not perfectly fit around the inflatable tube.
Therefore, the actuator’s motion under pneumatic pressurization can be divided into two phases:
the internal inflation phase and the deformation phase.

The internal inflation phase is the period during which the internal inflatable tube expands until it completely fills the space inside the sleeve formed by the embroidery and the fabric.
During this phase, almost no expansion force is transmitted to the exterior, and thus no deformation of the fabric surface occurs.
The inflation of the tube in this phase depends solely on the material properties of the inflatable tube and the applied pressure, and is independent of the embroidery design parameters.

Once the applied pressure reaches a certain threshold and the tube radius expands sufficiently to fill the internal cavity and contact the sleeve, the actuator transitions to the deformation phase.

In the deformation phase, further pressurization causes the inflatable tube to expand against the constraints of the embroidery and fabric, resulting in deformation of the entire fabric structure. 
The uniform fabric side provides a stronger constraint against expansion; therefore, the central axial length of the outer wall of the tube in contact with the fabric remains almost unchanged from its initial value.
Conversely, the embroidery-thread side exerts varying constraints depending on the embroidery pattern.
As a result, the central axial length of the region in contact with the embroidery changes according to the embroidery pattern.
The difference in axial length between the fabric side and the embroidery side generates bending deformation across the fabric surface.

In Section\ref{SSec_Model}, we analyze the actuator deformation in the deformation phase under applied pressure and construct an analytical deformation model.
In Section\ref{SSec_Interphase}, we describe the behavior of the inflatable tube during the internal inflation phase and define the transition point between the two phases.

To construct the analytical deformation model, the variables are defined as shown in Table~\ref{tab:variables}.
Among those variables, $l_0$, $r_f$, and $d_f$ are material parameters of the inflatable tube, and $\alpha_0$ and $w$ are design parameters defined by the embroidery pattern.

\begin{table}[t]
    \centering
    \caption{Symbols and definitions of variables.}
    \label{tab:variables}
    \begin{tabular}{c p{7cm}}
        Symbol & Definition\\
        \hline
        $l$ & The central axial length of the region of the inflatable tube in contact with the embroidery.\\
        \hline
        $l_0$ & The axial length of the inflatable tube in unpressurized state.\\
        \hline
        $r$ & The outer radius of the inflatable tube.\\
        \hline
        $r_f$ & The outer radius of the inflatable tube in unpressurized state.\\
        \hline
        $r_0$ & The outer radius of the inflatable tube at the moment the deformation phase begins.\\
        \hline
        $d$ & The inner radius of the inflatable tube.\\
        \hline
        $d_f$ & The inner radius of the inflatable tube in unpressurized state.\\
        \hline
        $P$ & The applied pneumatic pressure.\\
        \hline
        $P_0$ & The pneumatic pressure at the transition to the deformation phase.\\
        \hline
        $\alpha_0$ & The embroidery angle of the cross-pattern.\\
        \hline
        $w$ & The embroidery width.
    \end{tabular}
\end{table}

\subsection{Analyrical Deformation Model}
\label{SSec_Model}

We construct an analytical deformation model of the actuator after it comes into contact with the embroidery threads.  
In modeling the deformation of actuators that undergo constrained expansion of inflatable tubes, the \textit{Neo-Hookean model} and \textit{Lagrange's equation of motion} are often employed~\cite{Model_Hirei,Model1}.  
Accordingly, this study also adopts these formulations for modeling.  

The axial length \( l \) on the embroidery side is used as the generalized coordinate in the Lagrangian formulation.  
The motion is treated as a static model, and dynamic effects are neglected.  
Gravitational terms are also ignored.  
Based on these assumptions, we derive the relationship between the strain energy \( E_s \) and the generalized force \( F \).  
At the moment when the pneumatic pressure \( P_0 \) is applied and the inflatable tube first comes into contact with the embroidery threads, the strain energy and the pneumatic energy are in equilibrium.  
Therefore, the energy associated with the applied pressure \( P_0 \) is assumed to be entirely consumed in constraining the inflatable tube by the embroidery and does not contribute to the actuator deformation.

Using the \textit{Neo-Hookean model}, the strain energy \( E_s \) is expressed as follows:
\begin{equation}
    E_s=\int_V\frac{1}{2}G(I_1-3)dV
    \label{Eq_Ac_dV}
.\end{equation}
Here, the shear modulus \( G \) represents the overall mechanical property of the actuator, which behaves as an integrated structure of the inflatable tube and the sleeve.  
\( I_1 \) is expressed in terms of the axial, circumferential, and radial stretch ratios \( \lambda_1 \), \( \lambda_2 \), and \( \lambda_3 \) as follows:
\begin{equation}
    I_1=\lambda_1^2+\lambda_2^2+\lambda_3^2
    \label{Eq_Ac_I1}
.\end{equation}
Assuming that the actuator is made of an incompressible and isotropic material,  
the principal stretch ratios in the axial, circumferential, and radial directions, \( \lambda_1 \), \( \lambda_2 \), and \( \lambda_3 \), satisfy the following condition:
\begin{equation}
    \lambda_1 \lambda_2 \lambda_3=1
    \label{Eq_lam123=1}
.\end{equation}
The axial stretch ratio \( \lambda_1 \) at any point on the cross section of the inflatable tube is given by:
\begin{equation}
    \lambda_1=\frac{l+l_0}{2l_0 }+(l-l_0 )\frac{(d+a)sin\phi}{2rl_0 }      
.\end{equation}
Here, \( a \) denotes the depth within the wall thickness of the inflatable tube measured from the central axis,  
and \( \phi \) represents the rotational angle from the central axis.  
Each principal stretch ratio is considered with respect to its initial state, and, considering Eq.~(\ref{Eq_lam123=1}), the following condition must be satisfied:
\begin{equation}
    \lambda_2=\frac{r}{r_0} ,  \lambda_3=\frac{1}{\lambda_1 \lambda_2}
    \label{Eq_Ac_lam23}
.\end{equation}
By integrating Eq.~(\ref{Eq_Ac_dV}) in the cylindrical coordinate system and substituting Eqs.~(\ref{Eq_Ac_I1})–(\ref{Eq_Ac_lam23}), the following result is obtained:
\begin{dmath}
    E_s = \frac{1}{2}\pi G l_0 
    \left(
    \left(A^2+\frac{r^2}{r_0^2}-3\right)(r^2-d^2 )+ \frac{1}{4} B^2 (r^4-d^4 ) \\ +\frac{r_0^2}{r^2} \frac{2A}{B^2}  ((A^2-B^2 r^2 )^{-\frac{1}{2}}-(A^2-B^2 d^2 )^{-\frac{1}{2}})
    \right),
    \label{Eq_E_s0}
\end{dmath}
here, \( A \) and \( B \) are functions of \( l \) and are expressed as:
\begin{equation}
    A=\frac{l+l_0}{2l_0 },\      B=\frac{l-l_0}{2rl_0 }    
.\end{equation}
Because the elastomer is assumed to be incompressible,
 \( d \) can be approximated as:
\begin{equation}
    d=\sqrt{r^2 - \frac{r_f^2-d_f^2}{\lambda_m}},\ \ \ \lambda_m = \frac{l+l_0}{2l_0}
.\end{equation}

The generalized force \( F \) generated when the pneumatic pressure \( P \) is applied inside the actuator can be calculated based on the principle of virtual work:
\begin{equation}
    F=(P-P_0)\frac{\partial V_i}{\partial l}
,\end{equation}
here, \( V_i \) represents the internal volume of the inflatable tube and is given by:
\begin{equation}
    V_i = \frac{l+l_0}{2} \pi d^2
.\end{equation}

From the strain energy and the generalized force, \textit{Lagrange's equation} of static motion can be expressed as:
\begin{equation}
    L_f = -E_{s},\ -\frac{\partial L_f}{\partial l} = F 
    \label{Eq_Aigle}
.\end{equation}

Eq.~(\ref{Eq_E_s0})–(\ref{Eq_Aigle}) constitutes the fundamental equations of the analytical deformation model of the Embroidery Actuator.  
More specific analytical deformation models for each embroidery pattern can be derived by applying the geometric constraints introduced by the corresponding embroidery pattern to Eq.~(\ref{Eq_E_s0})–(\ref{Eq_Aigle}).

\subsubsection{Zigzag-pattern}
In the zigzag-pattern, the embroidery threads are almost perpendicular to the longitudinal axis of the inflatable tube.  
Therefore, it is assumed that little radial expansion occurs and only axial elongation takes place.  
Under this assumption, the following condition can be obtained:
\begin{equation}
    r=r_0
    \label{Eq_Zigzag}
.\end{equation}
By substituting Eq.~(\ref{Eq_Zigzag}) into the fundamental equations,  
the relationship between \( l \) and \( P \) for the zigzag-pattern Embroidery Actuator can be obtained.  
Furthermore, \( l \) can be expressed using the bending angle \( \theta \) as follows:
\begin{equation}
    l = -2r_0\theta+l_0
    \label{Eq_Z_l-theta}
.\end{equation}
By substituting Eq.~(\ref{Eq_Z_l-theta}) into the relationship between \( l \) and \( P \),  
the relationship between the applied pneumatic pressure \( P \) and the bending angle \( \theta \) can be obtained.

\subsubsection{Cross-pattern}
In the cross-pattern, when the embroidery threads are pushed apart in the radial direction,  
a contraction effect occurs due to the pantograph mechanism, which is also observed in McKibben-type artificial muscles.  
Let \( \beta_0 \) denote the braiding angle of the embroidery thread at the moment when the inflatable tube has a radius of \( r_0 \),  
and \( \beta \) denote the braiding angle when the radius is \( r \).  
It is assumed that the radius \( r \) and the axial length \( l \) satisfy the following relationship:
\begin{equation}
    \frac{l}{l_0}=\frac{\sin(\beta)}{\sin(\beta_0)},\ \ \ \frac{r}{r_0}=\frac{\cos(\beta)}{\cos(\beta_0)}
\label{Eq_Cross}
.\end{equation}
Here, based on geometric analysis, \( \beta_0 \) can be calculated from the design parameters as follows:
\begin{equation}
    \beta_0=\arcsin\left(\frac{(w\tan(\alpha_0))}{4r_f^2+\frac{w^2}{4}+\frac{w^2}{4}\tan^2(\alpha)}\right)
.\end{equation}
By substituting Eq.~(\ref{Eq_Cross}) into the fundamental equations,  
the relationship between \( l \) and \( P \) for the cross-pattern Embroidery Actuator can be obtained.  
Furthermore, \( l \) can be expressed using the bending angle \( \theta \) as follows:
\begin{equation}
    l = \frac{l_0\left(1-\theta\gamma\sqrt{\cos^2{(\beta_0)}+\theta^2 \gamma^2 \sin^2{(\beta_0)}}\right)}
    {1+\theta^2\gamma^2\sin^2{(\beta_0)}}
    \label{Eq_C_l-theta}
,\end{equation}
here, \( \gamma \) is given by
\begin{equation}
    \gamma = \frac{2r_0}{l_0\cos{(\beta_0)}}
.\end{equation}
By substituting Eq.~(\ref{Eq_C_l-theta}) into the relationship between \( l \) and \( P \),  
the relationship between the applied pneumatic pressure \( P \) and the bending angle \( \theta \) can be obtained.

\subsection{Internal Inflating Phase}
\label{SSec_Interphase}
In this section, we derive the pneumatic pressure \( P_0 \) at which the outer radius \( r \) of the inflatable tube equals the internal cavity radius \( r_0 \),  
which represents the transition point of the analytical deformation model of the Embroidery Actuator.  

The radius \( r_0 \) can be expressed as follows, assuming that during fabrication,  
the fabric and embroidery threads form an isosceles triangle whose base is the embroidery width \( w \) and whose height is the initial outer diameter of the inflatable tube \( 2r_f \),  
and that the circumference of a circle with radius \( r_0 \) coincides with the perimeter of this isosceles triangle:
\begin{equation}
    r_0 = \frac{2\sqrt{(\frac{w}{2})^2 + (2r_f^2)}+w}{2\pi}
    \label{Eq_r0}
.\end{equation}

The relationship between \( r \) and \( P \) can be expressed using the \textit{Neo-Hookean model} and \textit{Ogden models},  
which relate the strain energy of the inflatable tube to the applied pneumatic energy.  
According to the Neo-Hookean , the strain energy \( E \) per unit volume of the expanding inflatable tube is given by:
\begin{equation}
    E=\frac{1}{2} G_e\left( 
    I_1-3
    \right)
    \label{Eq_Gomuhazime}
.\end{equation}
Here, \( G_e \) is the shear modulus of the rubber and is a parameter determined by the material properties of the inflatable tube.  
In addition, \( I_1 \) is the first invariant in the axial direction and can be expressed using the axial, circumferential, and radial strains \( \lambda_1 \), \( \lambda_2 \), and \( \lambda_3 \) as follows:
\begin{equation}
    I_1=\lambda_1^2+\lambda_2^2+\lambda_3^2    
.\end{equation}
If the axial elongation is assumed to be negligible, the characteristics of the elastomer lead to the following relationship among \( \lambda_1 \), \( \lambda_2 \), and \( \lambda_3 \):
\begin{equation}
    \lambda_1=1, \lambda_2=\frac{r}{r_f} ,\lambda_3=\frac{1}{\lambda_2}    
.\end{equation}
In addition, the radial stress \( \sigma_2 \) and the thickness-direction stress \( \sigma_3 \) of the inflatable tube satisfy the following relationship~\cite{Ogden2017}:
\begin{equation}
    \sigma_2-\sigma_3=\lambda_2\frac{\partial E}{\partial\lambda_2},\,\frac{\partial\sigma_3}{\partial r}+\frac{\sigma_3-\sigma_2}{r}=0
.\end{equation}
Here, if \( \sigma_3 \) is regarded as a function of the variable \( r' \), which represents the outer radius of the inflatable tube,  
the boundary conditions are given as follows:
\begin{equation}
    \sigma_3 = \left\{
    \begin{array}{l}
        P\ on\ r^\prime = r\\
        0\ \ on\ r^\prime = r_f
    \end{array}
    \right.
    \label{Eq_GomuKyokai}
.\end{equation}
From Eqs.~(\ref{Eq_Gomuhazime})–(\ref{Eq_GomuKyokai}), the relationship between \( P \) and \( r \) can be derived.
\begin{equation}
    P=\frac{1}{2}G_e\left(
    \frac{r^2}{r_f^2}+\frac{r_f^2}{r^2}-2
    \right)
    \label{Eq_r}
.\end{equation}

By substituting the obtained \( G \) from Eq.~(\ref{Eq_r}) and \( r_0 \) from Eq.~(\ref{Eq_r0}),  
the pneumatic pressure \( P_0 \) at which the phase transition occurs can be obtained.
As an example, in the calculated results, when the embroidery widths \( w \) are \( 5~\mathrm{mm}, 7~\mathrm{mm}, \text{and } 9~\mathrm{mm} \),  
the corresponding \( P_0 \) values are \( 25~\mathrm{kPa}, 85~\mathrm{kPa}, \text{and } 180~\mathrm{kPa} \), respectively.

\section{Experiment}
\label{Sec_Exp}

In this section, prototype Embroidery Actuators with zigzag and cross-patterns, as shown in Fig.~\ref{Fig_Prototype},  
were fabricated with various design parameters.  
Their deformation behaviors were investigated and compared with the analytical deformation model.

\subsection{Prototypes Design}

To validate the proposed analytical deformation model, seven Embroidery Actuators were fabricated using a natural-rubber inflatable tube (T-2 02-089-01-02, Hagitec), polyester embroidery thread, and cotton fabric as the base material. The mechanical properties of these components were kept identical across all samples to isolate the influence of embroidery geometry. Two types of embroidery patterns, zigzag-pattern and cross-pattern, were designed using embroidery software (DG16, TISM) and sewn using a composite fiber-sewing machine (TCWM-101, TISM).
For the zigzag-pattern actuators, the stitch interval was set to 1.0~mm, and the total actuator length \( l_0 \) was 100~mm.  Three embroidery widths $w$ were prepared: 5~mm, 7~mm, and 9~mm.
For the cross-pattern actuators, the stitch interval was set to 1.4~mm, the total actuator length \( l_0 \) was 100~mm, and the embroidery width was 7~mm. Four embroidery angles $\alpha_0$ were tested: 15~deg, 30~deg, 45~deg, and 60~deg.

\subsection{Experiment Protocol}
\begin{figure}[t]
    \centering
    \subfloat[]{\includegraphics[width=0.5\linewidth]{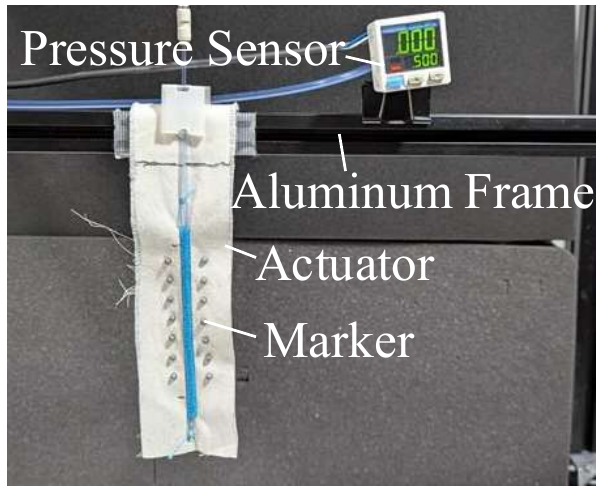}\label{Fig_Plat_fig}}
    \subfloat[]{\includegraphics[width=0.5\linewidth]{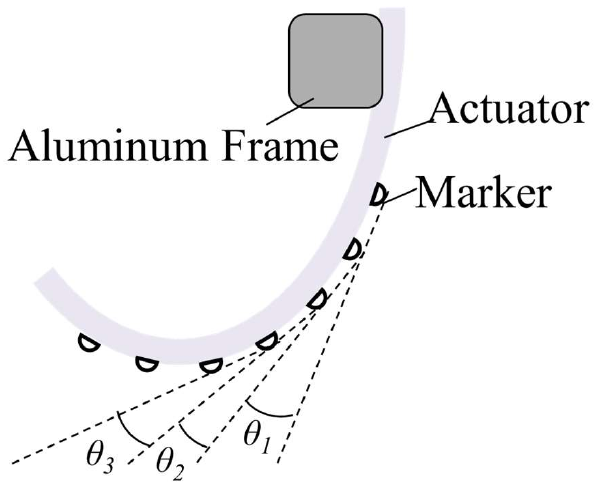}\label{Fig_Plat_eva}}
    \caption{Experiment platform. (a) 14 markers were attached to the prototype surface in two rows of seven at 10-mm intervals. The prototype was then suspended from the aluminum frame. (b) During bending, the angles (\( \theta_1, \theta_2, \ldots, \theta_5 \)) formed between the straight lines connecting adjacent markers were calculated, and the sum of the angles was defined as bending angle.
}
    \label{Fig_SensorRes}
\end{figure}

The experimental setup is shown in Fig.~\ref{Fig_SensorRes}.  
Each prototype was suspended vertically from an aluminum frame with the air-supply port facing upward.
The upper end of the fabric was fixed by a clip, and the lower end was free to deform. 
Pneumatic pressure was applied using an electro-pneumatic regulator (CRCB-0135W, Koganei), and the supplied pressure was measured by a pressure sensor (GS6, Koganei).
The deformation of each actuator was recorded using a motion-capture system equipped with seven infrared cameras (PrimeX 13,OptiTrack). 14 reflective markers were attached symmetrically on both sides of the embroidery region (seven per side). The three-dimensional coordinates of all markers were captured over time, and the bending angle was calculated as the sum of the local angles formed between adjacent marker segments, as shown in \ref{Fig_SensorRes}(b).
Pneumatic pressure was increased in increments of 10~kPa up to the maximum stable deformation for each sample. After each pressure increment, the actuator was held for 5~s to ensure quasi-static equilibrium before measurement. The same procedure was repeated during pressure release to examine hysteresis behavior.

\section{RESULTS AND DISCUSSION}
\label{Sec_Result}
\begin{figure}[t]
    \centering
    \subfloat[]{\includegraphics[width=0.32\linewidth]{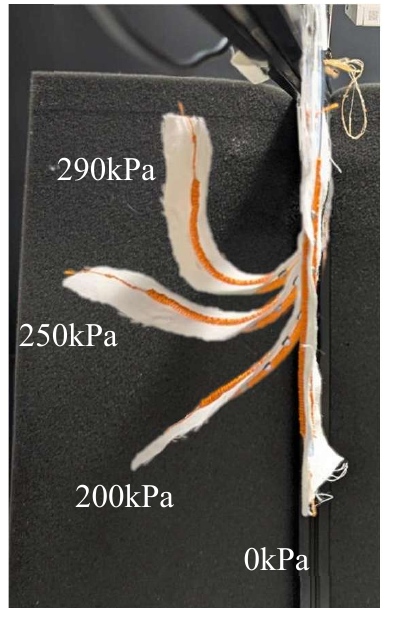}\label{Fig_def_z}}
    \subfloat[]{\includegraphics[width=0.32\linewidth]{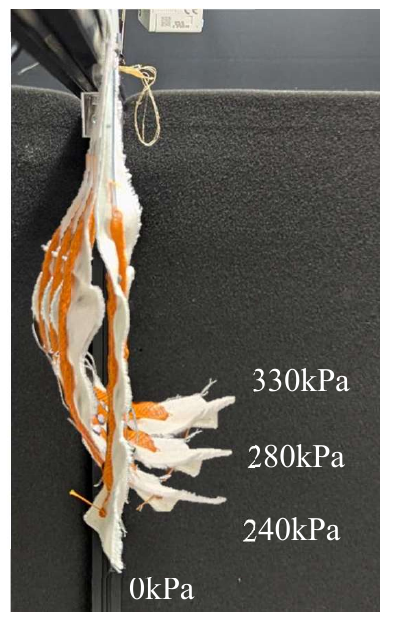}\label{Fig_def_c15}}
    \subfloat[]{\includegraphics[width=0.32\linewidth]{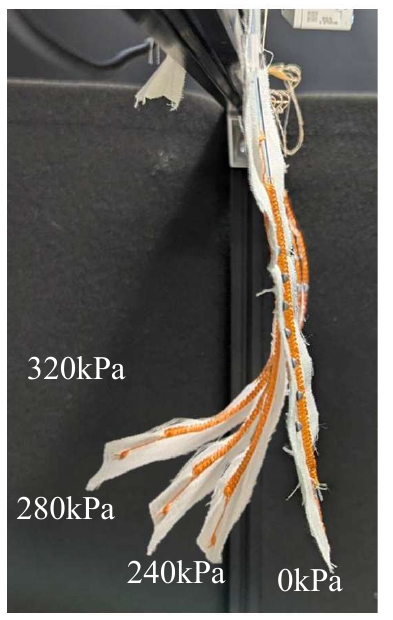}\label{Fig_def_c60}}
    \caption{The deformations of three representative patterns: (a)Zigzag-pattern actuator in witch $w$ is 7~mm bent toward the backside. (b)Cross-pattern actuator with $\alpha_0$ = 60~deg bending toward the frontside. (c)Cross-pattern actuator with $\alpha_0$ = 15~deg bending toward the backside. }
    \label{Fig_ExDeformation}
\end{figure}

.
The deformations of actuators with three representative patterns are shown in Fig.~\ref{Fig_ExDeformation}.
We discuss the experimental results and the corresponding analytical deformation models for the zigzag and cross-patterns.

\subsection{Zigzag-pattern}
\begin{figure}[t]
    \centering
    \includegraphics[width=1.0\linewidth]{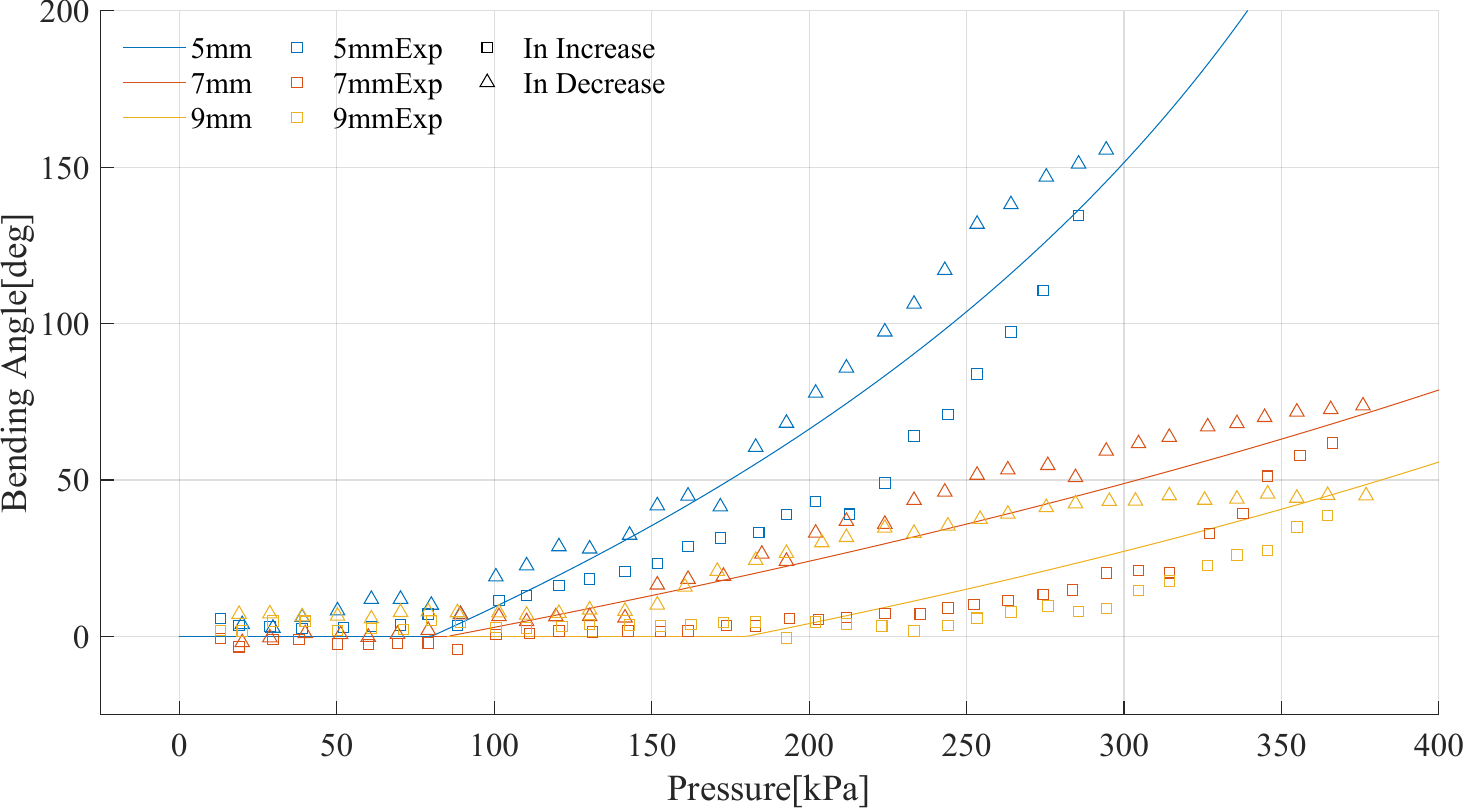}
    \caption{Analytical and experimental bending angles of zigzag-pattern. The solid lines represent the analytical deformation model plots, and the markers indicate the experimental data. The color indicates different embroidery widths, while the marker shape represents whether the pressure is increasing or decreasing.}
    \label{Fig_ZResult}
\end{figure}

The relationship between pneumatic pressure and bending angle of the zigzag-pattern actuator is shown in Fig.~\ref{Fig_ZResult}, and the model parameters are shown in TABLE~\ref{tab:zigzag}.
Across all embroidery widths, the actuator consistently bent toward the backside (positive direction) as shown in Fig.~\ref{Fig_ExDeformation}(a). As the embroidery width $w$ increased, the onset pressure required for deformation also increased—from approximately 50~kPa for $w$ = 5~mm to 180~kPa for $w$ = 9~mm.
 
The narrower embroidery widths produced larger bending under the same applied pressure, achieving a peak bending angle of approximately 155~deg for $w$ = 5~mm at 290~kPa.  
The bending–pressure curve exhibited nonlinear behavior, with the bending rate increasing sharply at higher pressures, and all samples showed noticeable hysteresis between pressurization and depressurization cycles.
The analytical deformation model demonstrated good agreement with the experimental results.The fitted shear modulus $G$ increased with embroidery width, indicating that larger widths generate stronger mechanical constraints between the fabric and the tube, thereby reducing extensibility. 
Minor discrepancies in $P_0$ observed at narrow embroidery widths can be attributed to the embroidery thread conforming more closely to the curved tube surface, effectively increasing the sleeve radius $r_0$.

\subsection{Cross-pattern}

\begin{figure}[t]
    \centering
    \includegraphics[width=1.0\linewidth]{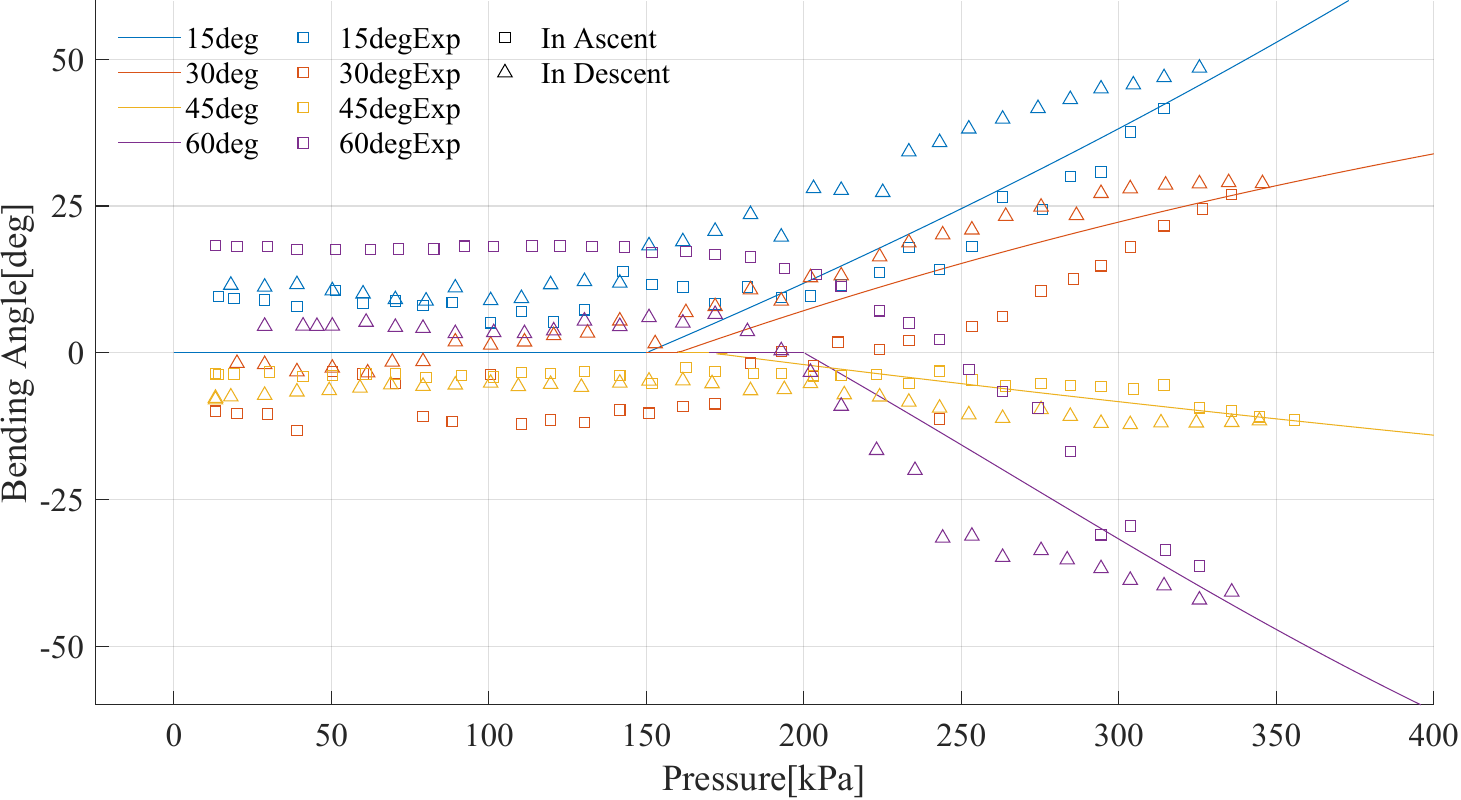}
    \caption{Analytical and experimental bending angles of cross-pattern. The solid lines represent the analytical deformation model plots, and the markers indicate the experimental data. The color indicates different embroidery angles, while the marker shape represents whether the pressure is increasing or decreasing.}
    \label{fig:CResult2}
\end{figure}

The relationship between pneumatic pressure and bending angle of the cross-pattern actuator is shown in Fig.~\ref{fig:CResult2}, and the model parameters are shown in TABLE~\ref{tab:cross}.

The deformation direction strongly depended on the embroidery angle: actuators with \( \alpha_0 \geq 45\)~deg bent toward the front side(negative side) as shown in Fig.~\ref{Fig_ExDeformation}(b), while those with \(\alpha_0 \leq 30\)~deg bent toward the back side(positive direction) as shown in Fig.~\ref{Fig_ExDeformation}(c).
For \( \alpha_0 = 60 \)~deg and 45~deg, bending was initiated at around 200~kPa, followed by rapid deformation. A peak bending angle of -48~deg was observed for the actuator with \( \alpha_0 = 60 \)~deg.
Conversely, actuators with $\alpha_0$ = 15~deg and  30~deg exhibited gradual bending in the positive direction. 
This trend suggests that, as the pantograph-induced contraction weakens, the axial extension of the inflatable tube becomes dominant, similar to the behavior of the zigzag-pattern actuator.
All samples exhibited hysteresis during pressurization–depressurization cycles. 
The experimentally determined transition pressure $P_0$ was generally higher than the predicted value, likely due to additional friction and local wrinkling introduced during the embroidery process. Nevertheless, the overall trends and magnitudes of bending closely matched the analytical deformation model, confirming its predictive validity.

\begin{table}[t]
    \centering
    \caption{Parameters in analytical bending angles of zigzag-pattern.}
    \label{tab:zigzag}
    \begin{tabular}{c||c|c|c}
        $w$[mm] & 5 & 7 & 9  \\
        \hline
        G[MPa] & 0.8 & 2.7 & 3.4 \\
        \hline
        $P_0$[kPa] & 50 & 85 & 180\\
    \end{tabular}
\end{table}

\begin{table}[t]
    \centering
    \caption{Parameters in analytical bending angles of Cross-pattern.}
    \label{tab:cross}
    \begin{tabular}{c||c|c|c|c}
        $\alpha_0$[deg] & 15 & 30 & 45 & 60 \\
        \hline
        G[MPa] & 2.9 & 12.0 & 1.3 & 2.9 \\
        \hline
        $P_0$[kPa] & 150 & 160 & 170 & 200\\
    \end{tabular}
\end{table}

\subsection{Discussion}

The experimental results validate the core hypothesis that the embroidery geometry governs the deformation characteristics of the proposed actuator. Specifically:

\begin{enumerate}
    \item Embroidery width $w$ controls the stiffness and onset pressure of bending.
    \item Embroidery angle $\alpha_0$ determines the bending direction and magnitude through a transition from fabric-constrained extension to pantograph-induced contraction.
    \item Both patterns exhibit a hysteresis characteristic of pneumatic soft actuators, mainly caused by the viscoelasticity of the tube material and the friction between the constituent materials.
\end{enumerate}

The close correspondence between analytical deformation models and experimental data demonstrates that the simplified geometric–mechanical model captures the essential deformation mechanics of the Embroidery Actuator. Moreover, the observed direction-reversible deformation, achieved purely by modifying embroidery geometry, shows that the proposed approach enables intuitive, design-driven control of fabric-based soft actuators without altering materials or fabrication processes.

\section{Conclusion}
\label{Sec_Conclusion}

In this paper, we present the Embroidery Actuator, a new type of fabric-integrated pneumatic actuator that realizes diverse deformation behaviors by controlling the geometry of embroidery patterns. Unlike previous fabric actuators that rely on embedded fiber actuators, the proposed method uses embroidery threads as structural constraints that convert the expansion of an inflatable tube into targeted fabric deformation.

Experiments using zigzag and cross embroidery patterns confirmed that the deformation direction and magnitude can be systematically controlled by geometric parameters such as embroidery width and embroidery angle. 
The analytical deformation model, developed based on the Neo-Hookean  and \textit{Lagrange's equation}, accurately predicted the observed pressure–bending relationships, achieving good agreement with experimental data.
These results demonstrate that the embroidery-based design approach enables high versatility and predictability in fabric actuator behavior.

Future work will focus on integrating conductive or sensing threads to develop multifunctional actuators, as well as exploring new embroidery geometries and material combinations to achieve richer and more complex deformation modes.

\section*{ACKNOWLEDGEMENTS}

This paper was also supported by TISM Co., Ltd. and Tajima Group. The authors gratefully acknowledge this support.

\vspace{11pt}

\bibliographystyle{IEEEtran}
\bibliography{bunken}

\vfill

\end{document}